\documentclass[journal=jacsat,manuscript=article]{achemso}
\usepackage[T1]{fontenc}
\usepackage[colorlinks=true,linkcolor=blue,citecolor=blue,urlcolor=blue]{hyperref}
\usepackage[utf8]{inputenc}
\usepackage{chemformula}
\usepackage{xcolor}
\usepackage{siunitx}
\usepackage[nameinlink,noabbrev,capitalize]{cleveref}

\title{Alkaline CO electro-oxidation:
Mechanistic Differences between Copper and Gold Single Crystals
and Peculiarities of various Copper Facets}
\author{Aarti Tiwari}
\affiliation{SurfCat, Department of Physics, Technical University of Denmark, 2800 Kgs. Lyngby, Denmark}
\alsoaffiliation{Interface Science Department, Fritz-Haber Institute of the Max Planck Society, 14195 Berlin, Germany}
\altaffiliation{These authors contributed equally to this work}
\author{Nitish Govindarajan}
\affiliation{Catalysis Theory Center, Department of Physics, Technical University of Denmark, 2800 Kgs. Lyngby, Denmark}
\altaffiliation{These authors contributed equally to this work}
\author{Hendrik H. Heenen}
\affiliation{Catalysis Theory Center, Department of Physics, Technical University of Denmark, 2800 Kgs. Lyngby, Denmark}
\author{Anton Bj{\o}rnlund}
\affiliation{SurfCat, Department of Physics, Technical University of Denmark, 2800 Kgs. Lyngby, Denmark}
\author{Karen Chan}
\affiliation{Catalysis Theory Center, Department of Physics, Technical University of Denmark, 2800 Kgs. Lyngby, Denmark}
\email{kchan@fysik.dtu.dk}
\author{Sebastian Horch}
\affiliation{SurfCat, Department of Physics, Technical University of Denmark, 2800 Kgs. Lyngby, Denmark}
\alsoaffiliation{Department of Engineering Technology, Technical University of Denmark, 2750 Ballerup, Denmark}
\email{shor@dtu.dk}
\date{July 2021}

\begin{document}

\maketitle
\begin{abstract}

 Understanding CO electro-oxidation is crucial towards designing catalysts for electrochemically oxidizing complex organic molecules. Earth-abundant Copper (Cu) has recently been demonstrated to exhibit high alkaline CO electro-oxidation activity, rivaling the previously acclaimed Gold (Au). Herein, we combine single crystal rotating disc electrode (RDE) experiments and \textit{ab initio} microkinetic modeling to understand the underlying reaction mechanisms on Cu and Au surfaces. Cu exhibits a facet-dependent activity with Cu(111) having a 0.27~V lower overpotential than Au(111) and a comparable CO oxidation current density. Using Koutecky-Levich analysis and DFT based microkinetic modeling, we identify the rate-limiting pathway to be Langmuir-Hinshelwood on Cu whereas Eley-Rideal on Au. We additionally present strikingly variant RDE responses on four Cu facets (111, 100, 110 and 211) and long-term stability analysis on Cu(111) and Au(111). We find a combined reset-reaction profile helps Cu retain its high activity and pose a strong competition to the expensive Au catalysts.

\end{abstract}
\vspace{0.5cm}
\noindent

The oxidation of CO to \ch{CO2} has been extensively studied on several catalysts both in the context of thermal and electro-catalysis.\cite{Freund2011,Rodriguez2012,Chen2018}
The electrochemical process (e-COOR) has both fundamental and technological relevance. For instance, it has been used to understand pH dependence in electrochemical reactions, \cite{Paramaconi2014} it occurs in the electro-oxidation of biomass derived molecules\cite{holewinski2019} and can be applied to remove CO impurities from reactant streams to avoid poisoning of electrodes used in direct alcohol fuel cells. \cite{Parsons1988,Rizo2021}. CO thus has an ambivalent role in electrocatalysis being a central intermediate in many reactions on the one hand and a disruptive contaminant on the other.

In this article, we focus on the e-COOR, where both the intricate adsorption dynamics of CO and its subsequent coupling with OH, leading to the oxidation of CO, have been identified as important aspects.\cite{Shiraishi2014,Chen2017}
These aspects have already been extensively investigated on Platinum (Pt)---one of the most active monometallic electrocatalysts for fuel cell applications \cite{Petrii2019}.
Pt is prone to poisoning by CO via the blocking of active sites upon its exposure either as a reaction intermediate or an adventitious entity, for example during electrocatalytic operation.\cite{Shiraishi2014} The poisoning effect of CO on Pt has been the focus of research for decades\cite{Valdes-Lopez2020}, and CO tolerant alternatives including bimetallic alloys of Pt\cite{Liu2013} or other cost-effective elements have been reported.\cite{Liu2016} CO poisoning therefore necessitates electro-oxidation catalysts to also be capable of performing e-COOR in order to retain their activity. This approach has been exemplified by combining Pt with Ru where the latter supplies adsorbed *OH species to react with *CO species on the former to facilitate e-COOR \cite{Watanabe1975a,F.Lin1999} and also reactivates the poisoned Pt. \cite{Watanabe1975,Markovic1995} 

In the search for more CO-tolerant e-COOR catalysts, Au has hitherto been demonstrated as the most active catalyst in alkaline media, with an onset potential that is ca.\ 0.5 V lower than Pt. In addition, the e-COOR activity on Au is not short lived unlike Pt due to its weak interaction with CO, thereby avoiding surface poisoning. \cite{Rodriguez2010} Ever since the first observation of Au as an active e-COOR catalyst by Kita et al.,\cite{Kita1985} several studies aiming to understand the origin of the high activity of Au towards CO electro-oxidation have been reported. \cite{weaver1996,Rodriguez2010,Rodriguez2012} It is only recently that our work\cite{Tiwari2020} together with Kunze et al. \cite{Auer2020} has demonstrated earth abundant Cu to rival the activity of Au for alkaline CO electro-oxidation, thereby opening new avenues for this important reaction. The reason for the low overpotentials of Cu towards e-COOR has been attributed to the following experimental and theoretical findings:

\begin{enumerate}
    \item The ability of different Cu facets to have a significant *OH coverage \cite{Tiwari2020n} before the equilibrium potential for e-COOR (ca. -0.1 V vs. RHE)
    \item Favorable CO adsorption energetics on Cu surfaces 
    \item The co-existence of *CO and *OH in the relevant potential range to allow for the formation of the *COOH intermediate with surmountable barriers leading to the final oxidation product (\ch{CO2})
\end{enumerate} 

While these interesting findings have led to an initial understanding of the origin of the high e-COOR \emph{activity} on Cu, a pertinent question remains to be answered: how do the \emph{kinetics} of e-COOR on Cu compare to Au? To answer this, we present a combined theoretical-experimental study of e-COOR on well-defined Au and Cu single crystals (SCs). We first show the fundamental electrochemical response on Cu and Au SCs both in the absence and presence of CO, and then combine RDE experiments with density functional theory (DFT) based microkinetic modeling to evaluate the differences in their respective e-COOR mechanism. While we mostly concentrate on the (111) facet when comparing e-COOR, we also present the RDE response on other fundamental Cu facets, namely (100), (110) and (211), which turn out to be strikingly different. Finally, we explore and optimize the parameters to achieve long term stability (20 h) for alkaline e-COOR, which has rarely been pursued even for the well studied Au electrodes.


\begin{figure*}[hbt!]
    \centering
    \includegraphics[width=0.85\textwidth]{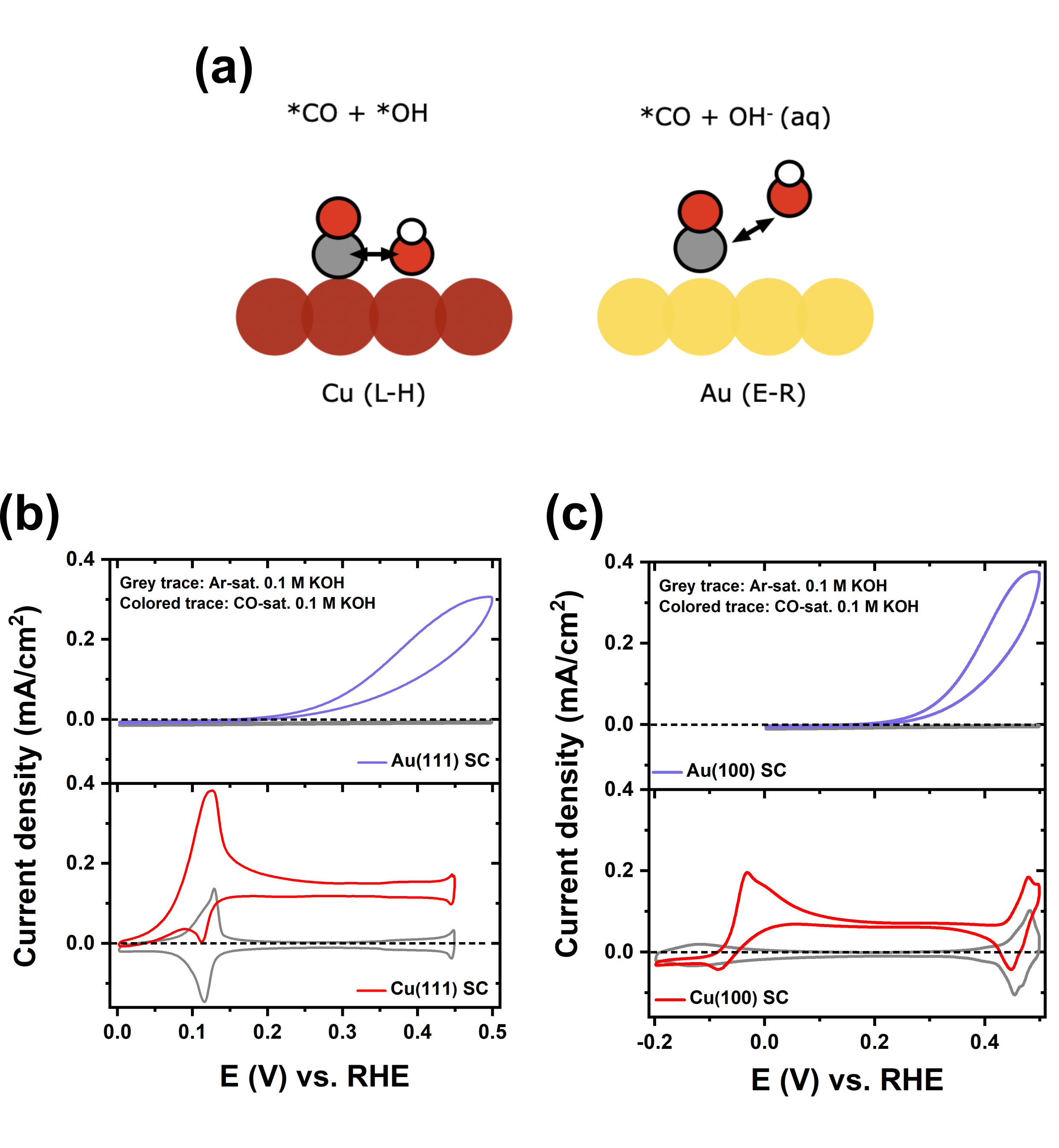}
    \caption{(a) Schematic for the proposed rate-determining step (RDS) for alkaline e-COOR on Cu and Au SCs involving the Langmuir-Hinshelwood (L-H) and Eley-Rideal (E-R) coupling pathways. Cyclic voltammograms for (b) Au \& Cu(111), and (c) Au \& Cu(100) SCs measured in blank Ar-saturated (grey trace) and CO-saturated (colored trace) 0.1~M KOH at a scan rate of 50~mV/s.}
    \label{fig:schematic-CV}
\end{figure*}

To set the stage, \cref{fig:schematic-CV}a presents the proposed rate-determining step (RDS) for alkaline e-COOR on Cu and Au SCs as identified from our rotating disc electrode (RDE) experiments and DFT based microkinetic modeling (\textit{vide infra}). As already suggested in previous studies on Cu SCs,\cite{Tiwari2020,Auer2020} our kinetic analysis confirms that the CO-OH coupling proceeds via a Langmuir-Hinshelwood (L-H) pathway (a chemical step), while it proceeds via an Eley-Rideal (E-R) pathway (an electrochemical step) on Au SCs, in agreement with previous studies on Au surfaces.\cite{weaver1996,Rodriguez2010}

The overall e-COOR is given by:
 \setcounter{equation}{0}
 \begin{align} 
 	& \mathrm{ CO + H_2O \rightarrow CO _2 + 2 H ^{+} + 2 e ^{-} \hspace{6em} (E{^0}= -0.10~V~ vs.\ RHE) }
 	 \label{e-COOR}
\end{align}

Proposed e-COOR mechanism on Cu:
\begin{align} 
 	& \mathrm{ CO (aq) + \ast \rightleftharpoons \ast CO } \\
 	& \mathrm{ H_2O(l) + \ast \rightleftharpoons \ast OH +   H ^{+}  + e ^{-} \hspace{6em} (x2) } \\
 	& \mathrm{ \ast CO +  \ast OH  \rightleftharpoons   \ast COOH +  \ast \hspace{7em} (RDS) } \label{RDS-Cu}  \\
 	& \mathrm{ \ast COOH +   \ast OH  \rightleftharpoons  CO _2 (aq) + H_2O (l) + 2 \ast } 
 	 \label{mechanism-Cu}
\end{align}

Proposed e-COOR mechanism on Au:
\begin{align}
 	& \mathrm{CO (aq) + \ast \rightleftharpoons \ast CO}  \\ 
 	& \mathrm{ \ast CO + OH^{-}(aq) \rightleftharpoons \ast COOH + e^{-} \qquad (RDS)} \label{RDS-Au} \\
 	& \mathrm{\ast COOH \rightleftharpoons CO_2(aq) + H^{+} + e^{-}}
\end{align} 

For e-COOR on Cu, the adsorption of the OH species (*OH) is the sole electrochemical step.  In contrast, on Au, both the E-R rate-determining step involving an OH$^{-}$ species and the formation of \ch{CO2} are electrochemical.\cite{Rodriguez2012}. We will now detail the experimental and computational kinetic studies that led us to the proposed rate-determining steps for alkaline e-COOR on Cu and Au SCs.

\begin{figure*}[htbp]
    \centering
    \includegraphics[width=0.9\textwidth]{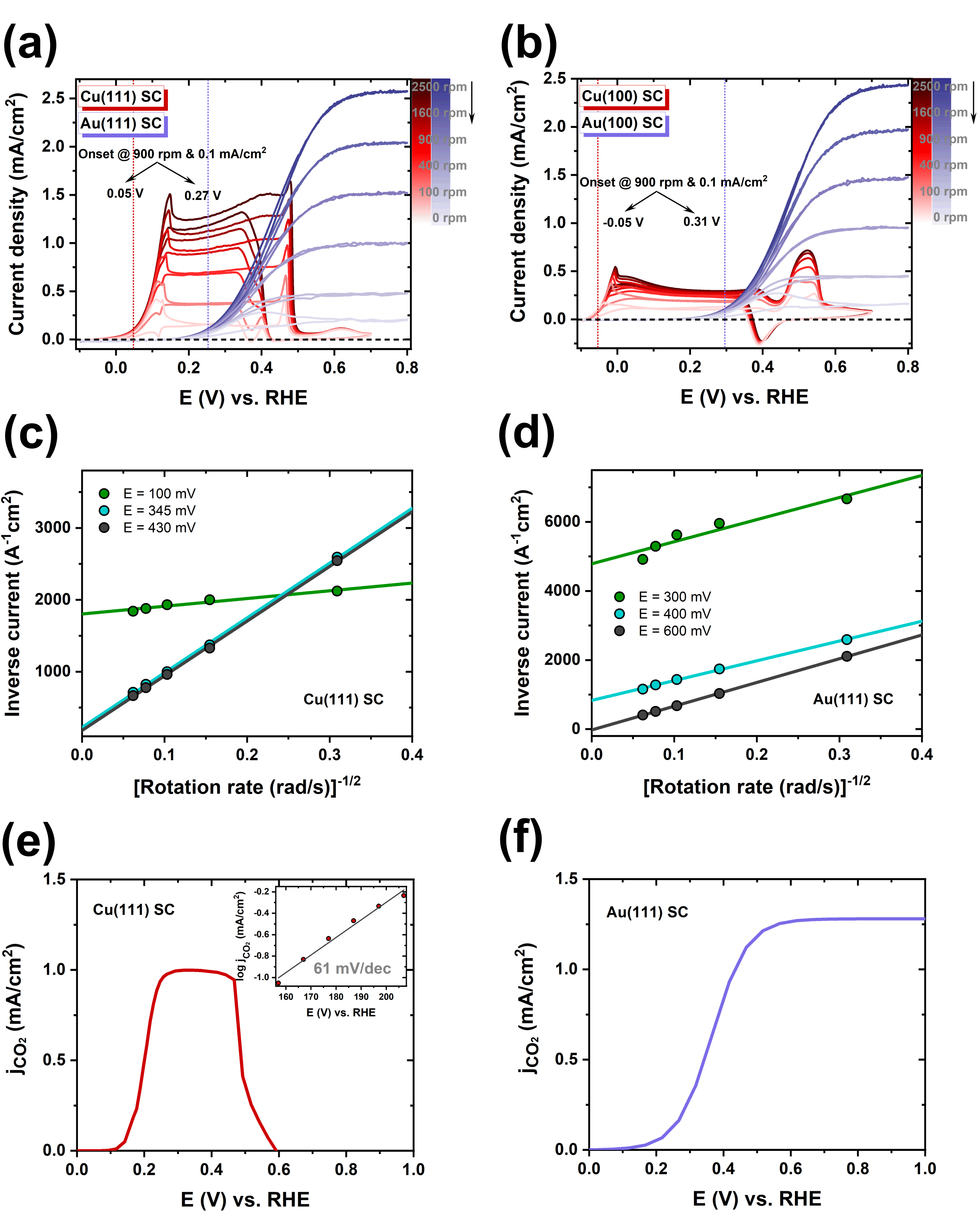}
    \caption{Polarization curves for Cu and Au SCs on the (a) (111) and, (b) (100) facets respectively in CO saturated 0.1~M KOH from 2500 to 100 rpm at a scan rate of 25~mV/s. Cu and Au are co-plotted here for easier comparison, in addition to a 0 rpm trace to visualize the effect of rotation. (c \& d) Koutecky-Levich plot for Cu(111) and Au(111) SCs respectively. Simulated CO$_2$ polarization curves for e-COOR on (e) Cu(111) with the simulated Tafel slope (61 mV/dec) close to the onset potential shown in the inset and (f) Au(111), both obtained using DFT based microkinetic modeling simulations.}
    \label{fig:Cu111 n Au111 comp}
\end{figure*}


Experimentally, we first measured cyclic voltammograms (CVs) on Cu and Au SCs to compare their respective e-COOR ability. \cref{fig:schematic-CV}b \& c show that for both Cu and Au SCs, the presence of CO(g) gives rise to a distinct oxidation current vs. the Ar-saturated case, which indicates the occurrence of e-COOR \cite{Tiwari2020,Auer2020} (cf.\ Section 1.1, SI for setup details \cite{Tiwari2019} and measurement parameters). Similar CV traces for Cu (110) and (211) are shown in Figure S1 (SI). We observe that CO electro-oxidation on both Cu(111) \& Cu(100) start at a potential closer to the equilibrium potential (ca. -0.1 V vs.\ RHE) compared to the respective Au facets (\cref{fig:schematic-CV}b \& c), which suggest Cu to be a relatively better e-COOR catalyst than Au. The respective onset and overpotentials ($\eta$) for Cu and Au SCs are provided in \cref{Onset table} and Table S1 (SI). To further evaluate the impact of mass transport limitations and understand the reaction kinetics, we performed RDE measurements. \cref{fig:Cu111 n Au111 comp}a \& b show a direct comparison between the e-COOR polarization curves for the (100) and (111) facets of Cu and Au. At the outset, Au SCs reveal a typical sigmoidal behavior with negligible hysteresis. On the contrary, Cu facets have strikingly different polarization curves with a distinct hysteresis suggesting that the steady-state coverage of adsorbates vary on a reduced vs. partially oxidized surface.\cite{WIBERG2012262}

\begin{table}[hbt!]
\centering
\caption{Experimentally observed onset ($E\textsubscript{onset}$) and overpotential ($\eta$) corresponding to $\SI{1}{\milli \ampere / cm^2}$ along with the diffusion limited current density at 900 rpm (\emph{j}) taken at 0.3~V for Cu SC and 0.6~V for Au SC from \cref{fig:Cu111 n Au111 comp}a \& b.}
\label{Onset table}
\begin{tabular}{lcccl}
\hline
$Surface$ & $E\textsubscript{onset} (V)$ & $\eta (V)$ & $\emph{j} (mA/cm^2)$ & \\ \hline
Cu(111) & 0.05                                        & 0.15                    & 0.97                                               &  \\
Au(111) & 0.27                                        & 0.37                    & 1.48                                               &  \\
Cu(100) & -0.05                                       & 0.05                    & 0.27                                               &  \\
Au(100) & 0.31                                        & 0.41                    & 1.39                                               &  \\ \hline 
\end{tabular}
\end{table}

Notable quantitative e-COOR outcomes from \cref{fig:Cu111 n Au111 comp}a \& b are given in \cref{Onset table} when compared at 900~rpm. Cu(111) exhibits superior e-COOR onset potential with an $\eta$ $\approx$220~mV lower than Au(111), while the current density (\emph{j}) on Au(111) is ca. 1.5 times higher than Cu(111). On the other hand, the onset potential of Cu(100) is ca. 360~mV  lower than Au(100), albeit with a significantly lower \emph{j}. Hence, both Cu SCs offer low overpotentials towards e-COOR and both Au SCs offer higher \emph{j}. However, considering both the aspects of onset and \emph{j} an optimal activity is observed over Cu(111).

The RDE results from \cref{fig:Cu111 n Au111 comp}a were further analyzed to extract the kinetics and mechanistic information for alkaline e-COOR on the (111) facet of both Cu and Au. Firstly, Koutecky-Levich plots shown in \cref{fig:Cu111 n Au111 comp}c,d were constructed using the Koutecky-Levich equation (cf. \cref{koutecky-levich}).

 \begin{align} 
 	& \mathrm{\frac{1}{j_L} = \frac{1}{j_K} + \frac{1}{j_D}} \label{koutecky-levich} \\ 
 	& \mathrm{j_D \propto {\omega}^{-1/2}}
 	\label{jD-omega}
\end{align}

Here, j$_L$ is the limiting current density, j$_K$ is the kinetic current density and j$_D$ is the diffusion-limited current density that is related to the rotation rate ($\omega$) (cf. \cref{jD-omega}). For Au(111), the Koutecky-Levich plot has a y-axis intercept $\approx$ 0 at 0.6 V, which suggests that only diffusion limitations govern the obtained current density at this potential. Conversely, Cu(111) has a non-zero intercept at ca. 0.4 V, indicating that even if the rotation rate is infinite, e-COOR will still have kinetic limitations for potentials $\geq$0.4 V. Further, j$_k$ in the absence of any mass transport limitations used to construct the Tafel plots (Figure 9, SI) are obtained from the Koutecky-Levich plot intercepts at different potentials. The estimated Tafel slope for Cu(111) is 70~mV/dec (Figure 9a, SI) suggesting the rate-limiting coupling of *CO and *OH after the first electron-transfer step \cref{RDS-Cu} (i.e.\ there is a Nernstian shift in activity w.r.t potential vs.\ SHE, corresponding to the change in *OH coverage with potential). In contrast, Au(111) exhibits a Tafel slope of 125~mV/dec (Figure 9b, SI), suggesting a rate-determining coupling of *CO and OH$^{-}$ in the first electron-transfer step (\cref{RDS-Au}) with a typical transfer coefficient between 0 and 1. The measured Tafel slope for Au(111) is in agreement with previous alkaline e-COOR studies.\cite{weaver1996,Rodriguez2010}

Additionally, DFT based microkinetic modeling was performed to obtain the simulated polarization curves for e-COOR on Cu(111) and Au(111). The details of the computational setup and simulations are provided in Section 2 (SI). The (chemical) association barriers of *CO and *COOH with *OH on Cu(111) was determined explicitly using DFT simulations (Figure S7, SI). Given the challenges associated with estimating electrochemical barriers including solvent effects and DFT simulations at a constant potential, \cite{Gauthier2019} we fit the barrier for the electrochemical coupling of *CO and OH$^{-}$ on Au(111) to match j$_k$ obtained using Levich analysis (cf. \cref{fig:Cu111 n Au111 comp}d). For this fit, the charge transfer co-efficient, $\beta$ for the CO-OH$^{-}$ coupling was assumed to be 0.5 (following the experimental Tafel slope of 125 mV/dec, Figure 9b, SI). Mass transport of CO(aq) was accounted with a Fickian diffusion model through a boundary layer of thickness $\approx$ 10 $\mu$m. We note that the boundary layer thickness has an effect on the magnitude of j$_D$ but the overall simulated polarization curve features remained unchanged (cf. Figure S10, SI). \cref{fig:Cu111 n Au111 comp}e \& f, shows the simulated  current density towards CO$_2$ (j$_{CO_2}$) on Cu(111) and Au(111) agreeing qualitatively with the RDE experiments (cf. \cref{fig:Cu111 n Au111 comp}a \& b). The simulated CO$_2$ polarization curves show e-COOR as entirely diffusion limited on Au(111) even at potentials $>$ 0.6 V, while Cu(111) is inactive. The latter arises from poisoning of Cu(111) due to the build-up of *OH species with increasing anodic potentials. Note that we do not consider surface oxidation reactions in our microkinetic model. Further, for Cu(111), we find the (chemical) coupling of *CO and *OH to be the rate-determining step close to the e-COOR onset potential as confirmed by a degree of rate control (DRC) analysis (Figure S11, SI).\cite{Campbell2009} This results in a simulated Tafel slope of 61 mV/dec (cf. \cref{fig:Cu111 n Au111 comp}e, inset) which agrees with the measured Tafel slope of 70 mV/dec (Figure 9a, SI).

\begin{figure*}[hbt!]
    \centering
    \includegraphics[width=0.9\textwidth]{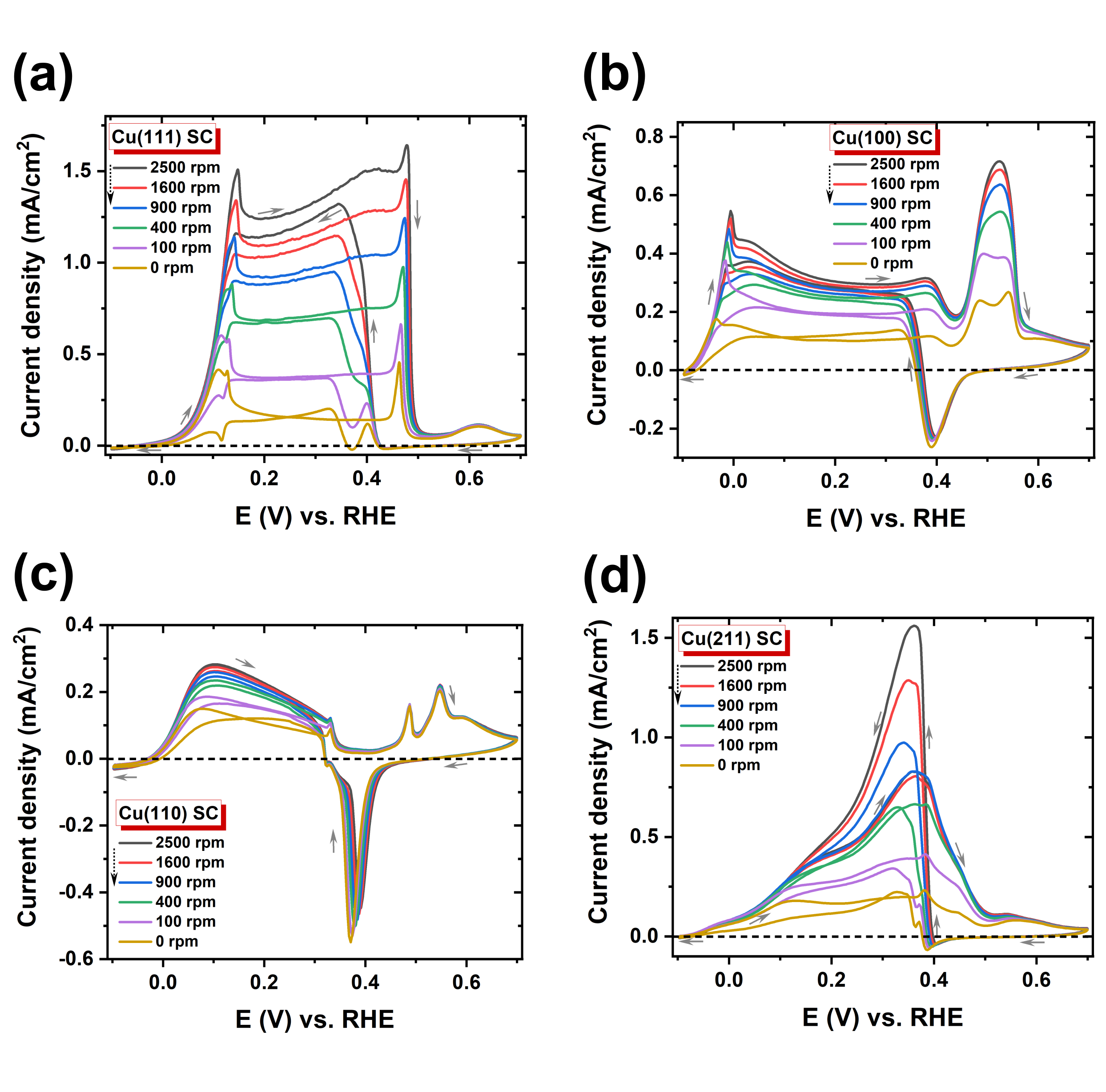}
    \caption{Polarization curves for (a) Cu(111), (b) Cu(100), (c) Cu(110), and (d) Cu(211) SCs in CO saturated 0.1~M KOH from 2500 to 0~rpm at a scan rate of 25~mV/s.
    Note the different y-axis.}
    \label{fig:RDE}
\end{figure*}

Having thus established the underlying differences between Cu and Au towards e-COOR, we now present the facet-specific RDE responses on Cu. \cref{fig:RDE} shows a complex facet-dependent RDE behaviour whose features as exemplified by the Cu(111) surface (\cref{fig:RDE}a) are interpreted as: (i) The first peak (at ca. 0.14~V) corresponds to OH adsorption on Cu(111). \cite{Maagaard2019,Tiwari2020} (ii) The second peak (at ca. 0.47~V) corresponds to the beginning of surface oxidation. \cite{Kunze-Liebhauser2018} (iii) The subsequent oxidative peak at ca. 0.6~V denotes progressive oxidation of Cu(0) to Cu(I) species \cite{Kunze2004} without any CO oxidation current [similar response to Ar-purged electrolyte; Figure S3a and S4a, SI]. (iv) On the reverse sweep, the surface first undergoes reduction (at ca. 0.4~V) and then regains its CO oxidation activity. (v) Compared to Au (cf. \cref{fig:Cu111 n Au111 comp}a \& b), a prominent hysteresis is observed between the forward and reverse trace on Cu(111). Similar characteristics, albeit with different features, are observed for Cu(100), Cu(110) and Cu(211) [cf. Section 1.2, SI]. Notably, enhanced mass transport only marginally increased the e-COOR current densities for Cu(110) whereas, Cu(211) exhibits relatively higher current densities on the reverse sweep (at ca. 0.34~V) suggesting that a freshly reduced surface is more active for e-COOR. A detailed investigation of these facet dependent features for e-COOR on Cu SCs warrants a separate study of its own.

\begin{figure*}
    \centering
    \includegraphics[width=0.9\textwidth]{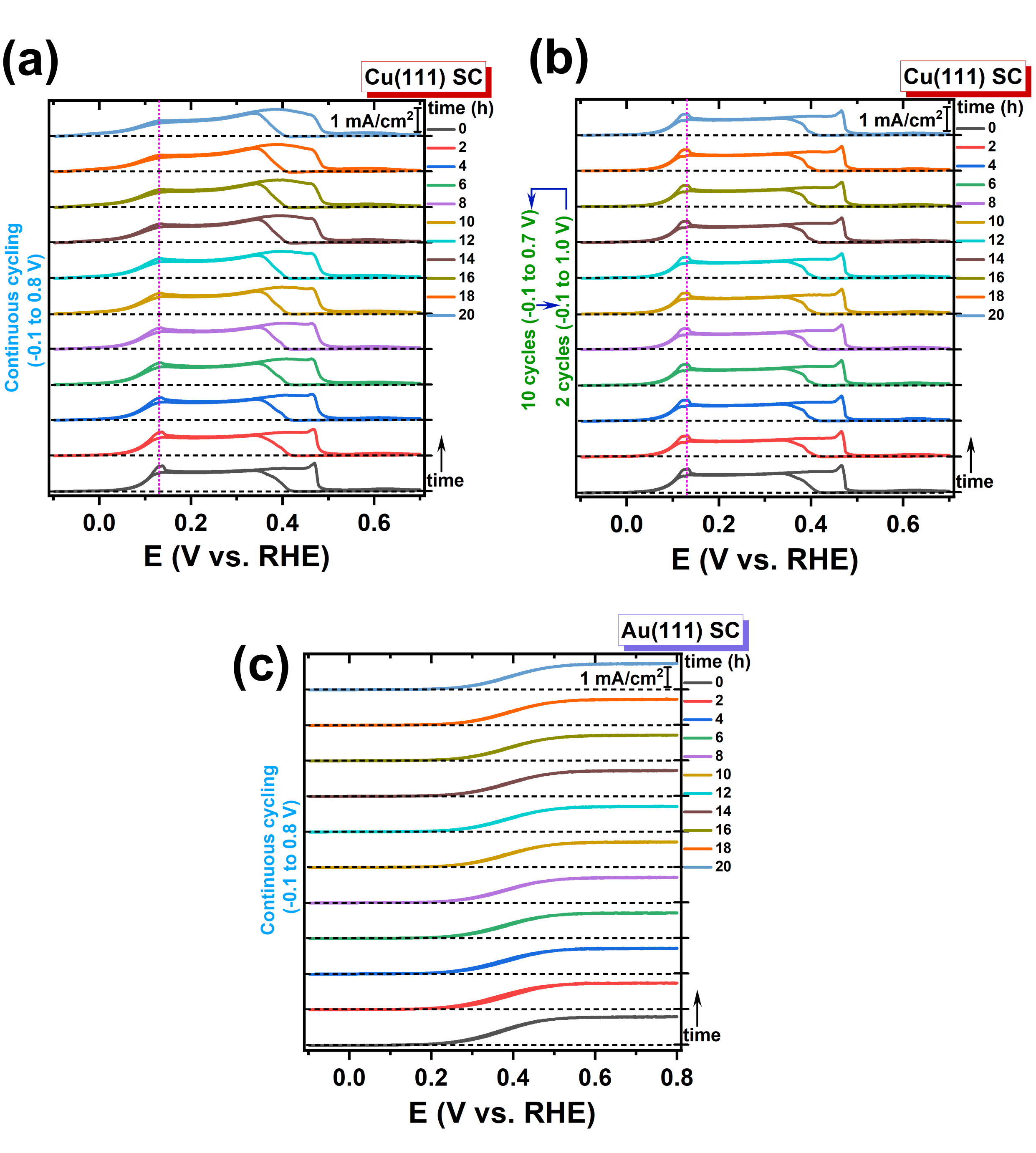}
    \caption{CO oxidation stability measurement under potentiodynamic conditions in the potential range of (a) -0.1 to +0.7~V for 20~h and (b) -0.1 to +0.7~V for 10 cycles followed by 2 cycles from -0.1 to +1.0~V for 2 cycles both repeated cyclically for 20~h on Cu(111); and (c) -0.1 to +1.0~V on Au(111) for 20~h in CO saturated (solid traces) and Ar saturated (dashed traces) 0.1~M KOH respectively at a scan rate of 25~mV/s.}
    \label{fig:Stability}
\end{figure*}

Finally, we address the long-term stability that, besides the electrocatalytic activity and reaction kinetics discussed above, is crucial to evaluate the overall performance of a catalyst. In the following, we focus on the stability of Cu(111) and Au(111) towards e-COOR. As seen from \cref{fig:Stability}a, even upon continuous cycling between -0.1 to +0.7~V, a significant CO oxidation activity on Cu(111) is retained for 20~h. However, Figure S5a (SI) clearly reflects progressive changes in the polarization curve with time, which could be attributed to variable coverage of the reactant, intermediates and product species on the catalyst surface, \cite{Tiwari2020} causing every potentiodynamic cycle to be affected by the history of the previous oxidative cycle. This aspect suggests if one could somehow reset the surface, it might be possible to retain the e-COOR activity. Considering these aspects and the fact that an oxidized Cu surface is inactive towards e-COOR but a freshly reduced surface is activated (\cref{fig:RDE}), another potential profile was designed. The first 10 cycles were measured between -0.1 to +0.7~V (reaction) followed by 2 cycles from -0.1 to +1.0~V (reset). A repetition of this specific combination of two potential profiles (reaction and reset) for 20~h lead to a long term retention of the CO oxidation activity for Cu(111) as shown in \cref{fig:Stability}b and Figure S5b (SI). In contrast, potentiodynamic cycling between -0.1 to +1.0~V for 20~h on Au(111) shows a relatively stable e-COOR response (\cref{fig:Stability}c and S5c, SI) without the need for a reset potential as compared to Cu(111). These differences can be attributed to the higher surface oxidation potential and weaker interaction of CO with Au(111) compared to Cu(111).

In summary, we investigated alkaline CO electro-oxidation on Cu and Au single crystals. Among the four Cu SCs explored, Cu(111) emerged as the best candidate having favorable thermodynamics i.e., an overpotential of only 150~mV (at 0.1~$mA/cm^2$) compared to 370~mV for Au(111) and 410~mV for Au(100), and an optimal kinetics i.e., a high e-COOR current density of 0.97~$mA/cm^2$ (at 900~rpm) that is well within an order of magnitude compared to Au SCs. On the basis of Koutecky-Levich analysis and DFT based microkinetic modeling, we find e-COOR on Cu(111) to follow a Langmuir–Hinshelwood type chemical *CO-*OH coupling, while on Au(111), it proceeds via an electrochemical Eley–Rideal type *CO-OH$^{-}$ coupling. In particular, moderate binding energy of CO, co-existence of *OH and *CO on Cu(111) at relevant potentials, and their facile coupling barrier results in the high activity of Cu surfaces towards alkaline e-COOR. A combined reset-reaction profile allowed us to obtain a highly reproducible and retained CO electro-oxidation activity on Cu(111) for 20~h, comparable to Au(111). The details of the presented work, provide an in-depth comparison of the activity and stability of earth abundant Cu to the well-known Au electrocatalyst for alkaline e-COOR. These insights have potential applicability towards electrosynthesis of industrially important organic molecules and fuel cell applications.

\vspace{0.5cm}
\noindent

\section{Code \& Data availability}

The CatMAP input files, the DFT opitimized structures and NEB trajectories, and the transport model used for the microkinetic simulations will be made available on the   \href{https://github.com/CatTheoryDTU}{CatTheory} Github account upon publication.

\section{Acknowledgements}

The computational work was supported by a research grant (29450) from VILLUM FONDEN.   The experimental work by European Union’s Horizon 2020 research and innovation programme under the Marie Sklodowska-Curie
grant agreement no. 713683.

\suppinfo
Additional details on the experiments, DFT simulations and microkinetic model.

\bibliography{RDE_ref}
\end{document}


\clearpage
\section{Experimental CVs and polarization curves}

\subsection{Experimental Methods}

The electrochemical cell used for the measurements is a one compartment, 3-electrode PTFE cell (Pine Research, detailed elsewhere \cite{Tiwari2019,Tiwari2020n}) wrapped in an Ar-filled air bag. The Cu and Au SC working electrodes [Mateck, Jülich (DE); purity 99.9999\%; typical diameter of exposed surface is 6 mm; thickness 3 mm] were housed in a custom built PTFE-holder compatible with the Pine Research rotator's RDE shaft \cite{Tiwari2019}. The counter electrode employed was a Ni mesh (10.5~$cm^2$) and all the potentials were referenced against a reversible hydrogen electrode (RHE). All the electrochemical measurements were performed by a Bio-Logic SP200 potentiostat run by Bio-Logic’s EC-Lab software. Prior to measurements, the Cu SCs were electropolished for 60~s in a 66\% \ch{H3PO4} solution (85\% EMSURE, Merck) at 2.0~V (vs. Cu wire; Goodfellow, purity 99.9\%) whereas Au SCs were first flame annealed and then electropolished for 60~s at 2.4~V (vs. Au wire; Goodfellow, purity 99.99+\%) followed by a thorough rinsing with Millipore water (\SI{18.2}{M\ohm \cdot cm}). The electrolyte was a 0.1~M KOH solution (99.995 Suprapur, Merck) prepared in a PFA volumetric flask (Corning Life Sciences) using Millipore water. The gases used for purging were Ar (6.0, Air Liquide) for blank measurements as well as in the air bag, and CO (5.0, Air Liquide) passed through a carbonyl trap (Leiden Probe Microscopy) for the CO oxidation measurements. It should be noted here that high purity gases were found essential to obtain the long-term CO oxidation response observed. The cleanliness state of the measurements was continuously monitored by performing ICP analysis of the electrolyte and \emph{ex situ} XPS analysis of the SCs to rule out any trace contamination, especially due to the presence of metallic impurities. The reproducibility of the measurements was ascertained by repeating them at least three times. All the presented data in the article are as obtained and without any baseline correction.

The upper limit in potential for the Cu CVs were chosen to incorporate OH adsorption features but to avoid surface oxidation, while the lower limit on potential was chosen to  avoid hydrogen evolution reaction (HER).  These bounds define the fingerprint region that identifies the surface orientation of Cu\cite{Tiwari2020n}. The CVs for Au were also measured in a similar range for better comparison with Cu. However, a full range CV (from 0 V to 1.5 V vs. RHE) for Au is shown in Figure S2 (SI) to compare with previous reports.

\begin{figure*}[hbt!]
    \centering
    \includegraphics[width=1\textwidth]{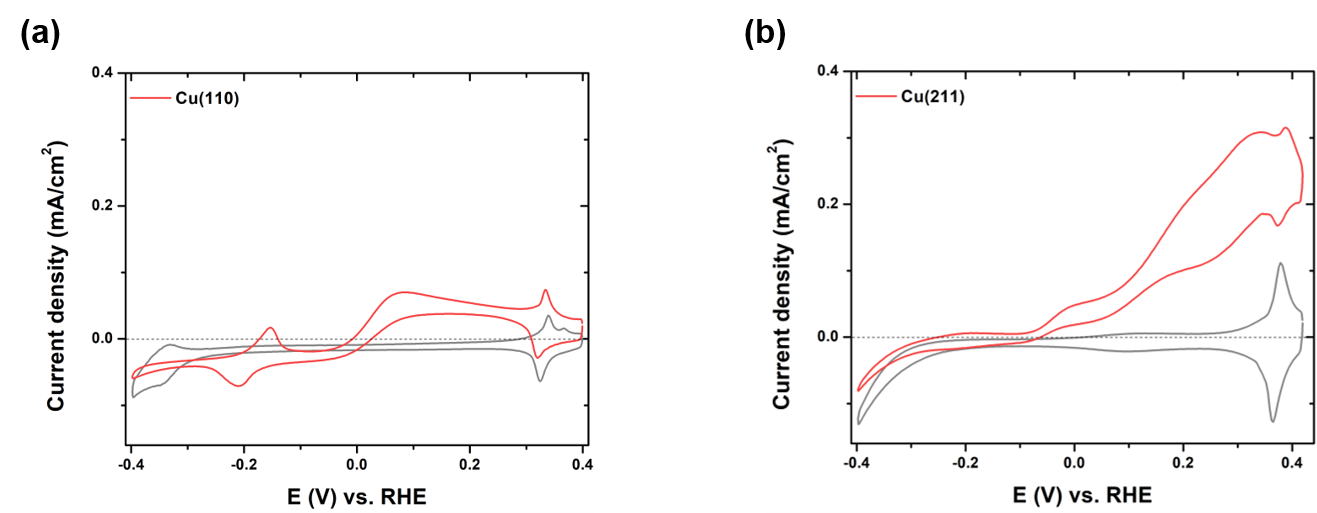}
    \caption{Cyclic voltammograms for (a) Cu(110), and (b) Cu(211) single crystals in their respective fingerprint regions measured in blank Ar-saturated (grey trace) and CO-saturated (red trace) 0.1~M KOH at a scan rate of 50~mV/s.}
    \label{fig:110 & 211}
\end{figure*}

\begin{figure*}[hbt!]
    \centering
    \includegraphics[width=1\textwidth]{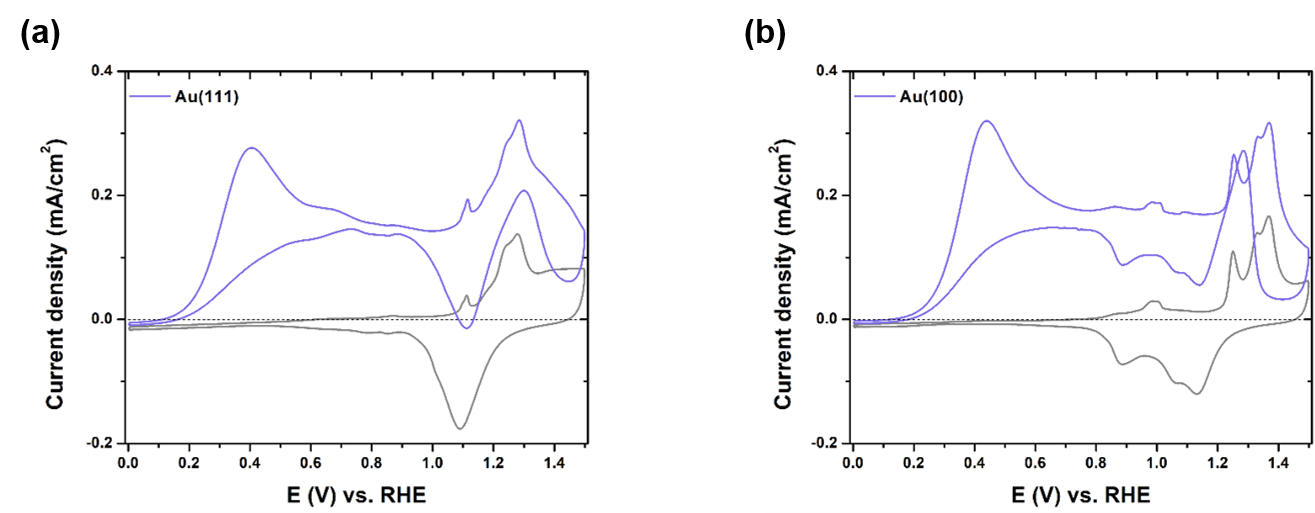}
    \caption{Cyclic voltammograms for (a) Au(111), and (b) Au(100) single crystals measured in blank Ar-saturated (grey trace) and CO-saturated (red trace) 0.1~M KOH at a scan rate of 50~mV/s.}
    \label{fig:Au long CV}
\end{figure*}

\begin{table}[bt!]
\centering
\caption{Experimentally observed onset ($E\textsubscript{onset}$) and overpotentials ($\eta$) corresponding to $\SI{25}{\micro \ampere / cm^2}$ obtained from the cyclic voltammograms in Figure 1 b \& c (Main text) and Figure S1 \& S2 (SI). The choice of current density i.e. $\SI{25}{\micro \ampere / cm^2}$ instead of $\SI{1}{\milli \ampere / cm^2}$ used in the main text is due to the low obtainable currents on the Cu(110) surface.}
\label{Onset table}
\vspace*{3mm}
\begin{tabular}{lccll}
\hline
$Surface$ & $E\textsubscript{onset} (V)$ & $\eta (V)$ &  &  \\ \hline
Au(111) & 0.25         & 0.35     &  &  \\
Au(100) & 0.28         & 0.38     &  &  \\
Cu(111) & 0.04         & 0.14     &  &  \\
Cu(100) & -0.07        & 0.03     &  &  \\
Cu(110) & 0.02         & 0.12     &  &  \\
Cu(211) & -0.05        & 0.05     &  &  \\ \hline
\end{tabular}
\end{table}

\begin{figure*}[hbt!]
    \centering
    \includegraphics[width=1\textwidth]{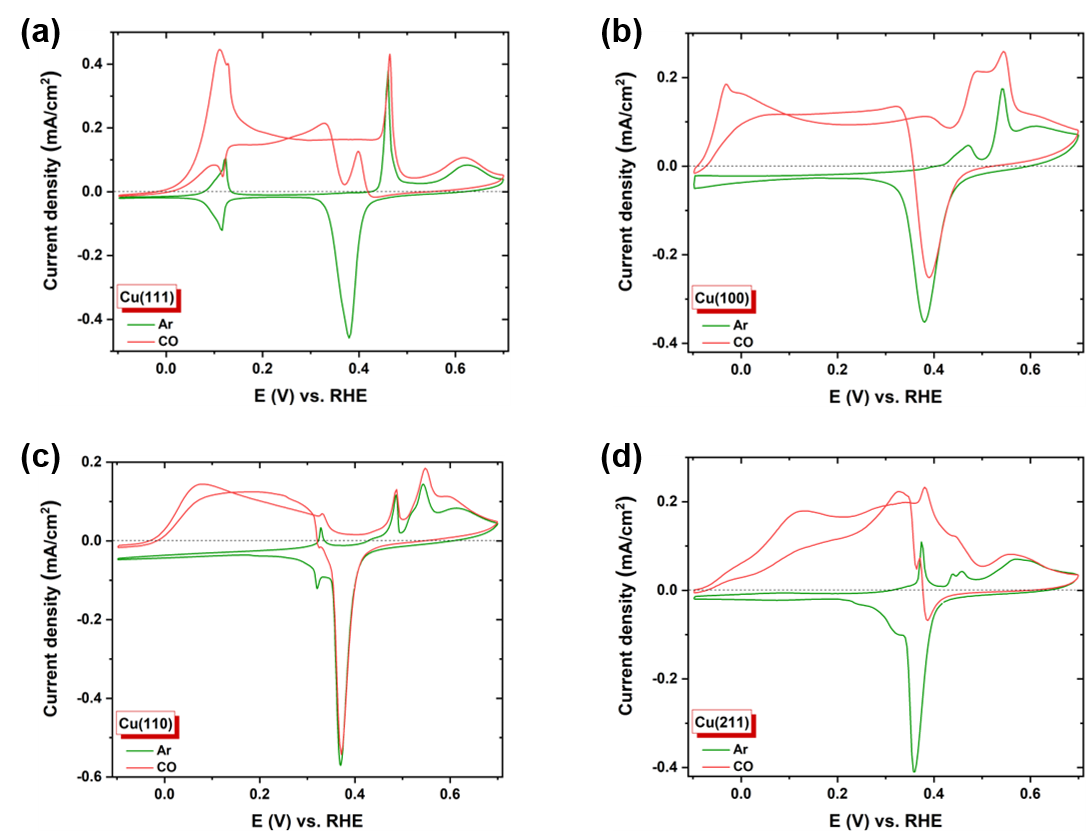}
    \caption{Polarization curves at 0~rpm for (a) Cu(111), (b) Cu(100), (c) Cu(110), and (d) Cu(211) single crystals in Ar (green trace) and CO (red trace) saturated 0.1~M KOH at a scan rate of 25~mV/s. The observed increase in oxidation currents under static conditions (i.e. 0 rpm) in the presence of CO vs. Ar suggests the respective abilities of the different Cu facets towards e-COOR. }
    \label{fig:Ar-CO-0rpm}
\end{figure*}

\begin{figure*}[hbt!]
    \centering
    \includegraphics[width=1\textwidth]{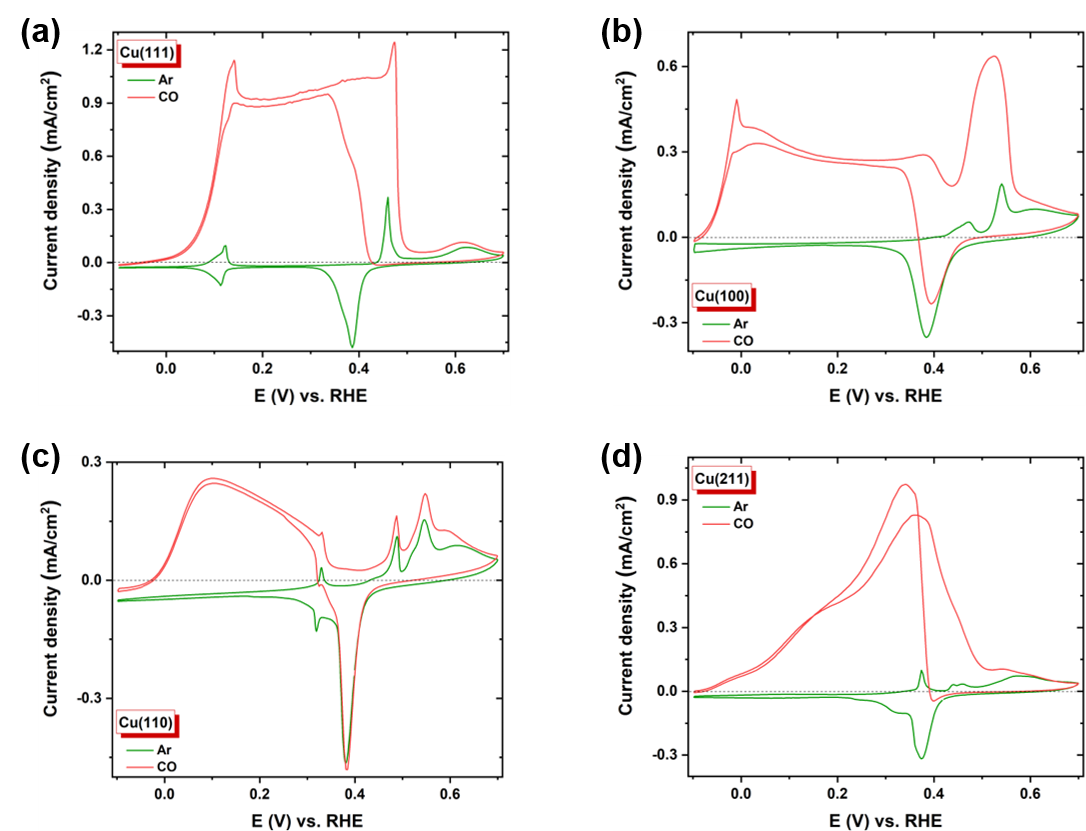}
    \caption{Polarization curves at 900~rpm for (a) Cu(111), (b) Cu(100), (c) Cu(110), and (d) Cu(211) single crystals in Ar (green trace) and CO (red trace) saturated 0.1~M KOH at a scan rate of 25~mV/s. A significant increase in oxidation current density compared to \cref{fig:Ar-CO-0rpm} depicts the influence of mass transport i.e. at 900~rpm (vs. 0 rpm) towards the respective Cu SCs e-COOR activity.}
    \label{fig:Ar-CO-900rpm}
\end{figure*}

\clearpage
\subsection{Details of RDE Features}
Following the discussion of Cu(111) RDE response (main text), Cu(100) also shows an initial increase in CO oxidation current at ca.\ 0~V (\cref{fig:Ar-CO-900rpm}b), which is consistent with the observation from CV measurements (Figure 1c, main text). It is followed by the limiting current behavior before merging with the surface oxidation response at ca.\ 0.38 and 0.52~V, after which the CO oxidation current dies off as the surface is oxidized. The renewal in activity post surface reduction and a slight hysteresis is a similar observation to Cu(111) but with lower e-COOR current densities. Cu(110) SC also shows a similar trend but there is only a marginal influence of enhanced mass transport on the obtained current density (Figure 2c, main text). This observation suggests the significant influence of a lower *OH coverage and more importantly of the surface structure that results in the presence of *OH and *CO in close proximity enabling CO electro-oxidation. \cite{Tiwari2020} Cu(211) also exhibits a similar response, but the only difference and an interesting observation is that the freshly reduced surface is highly active for CO oxidation. This characteristic is suggested by the higher obtainable oxidative current densities post surface reduction at ca.\ 0.34~V clearly seen upon enhanced mass transport of CO i.e., on increasing the rotation rate (rpm). In addition, the fact that CO is essential for the observed oxidative response is reflected clearly from \cref{fig:Ar-0 & 900rpm} depicting no difference between the RDE trace at both 0 and 900~rpm in blank Ar-saturated electrolyte for all the four Cu facets.

\begin{figure*}[hbt!]
    \centering
    \includegraphics[width=1\textwidth]{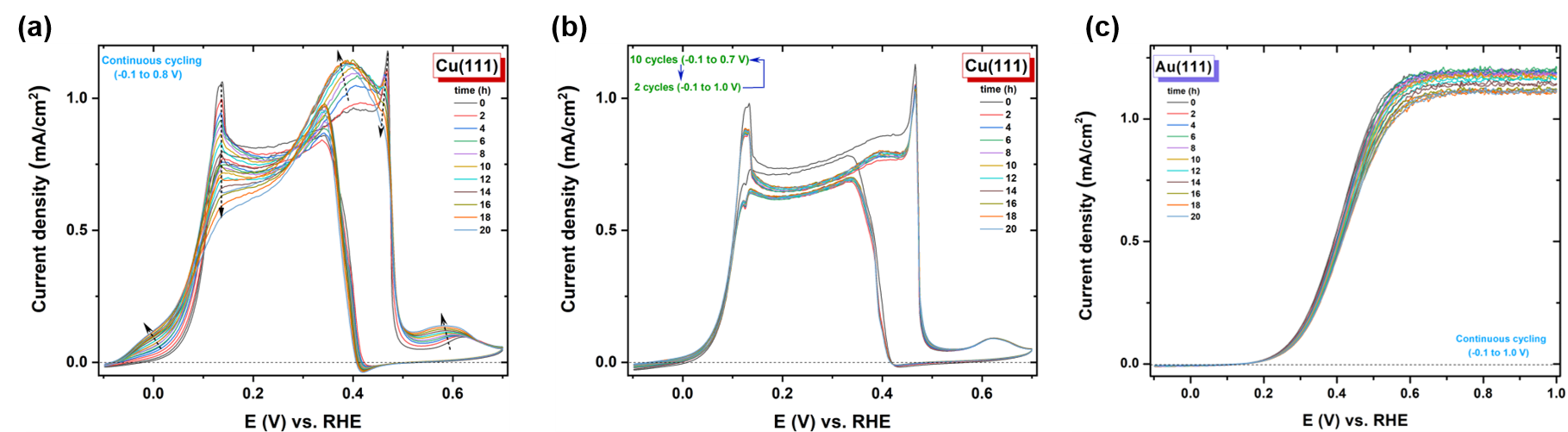}
    \caption{CO oxidation stability measurement under potentiodynamic conditions in the potential range of (a) -0.1 to +0.7~V for 20~h and (b) -0.1 to +0.7~V for 10 cycles (reaction profile) followed by from -0.1 to +1.0~V for 2 cycles (reset profile) both profiles repeated cyclically for 20~h on Cu(111); and (c) -0.1 to +1.0~V on Au(111) for 20~h in CO saturated 0.1~M KOH at a scan rate of 25~mV/s. The possibility of e-COOR activity retention for Cu(111) is depicted following a reset-reaction potential profile compared to a simple long term activity measurement. Similar measurements for Au(111) shows no need of such potential profiles and its retained long term activity.}
    \label{fig:Stability SI}
\end{figure*}

\begin{figure*}[hbt!]
    \centering
    \includegraphics[width=1\textwidth]{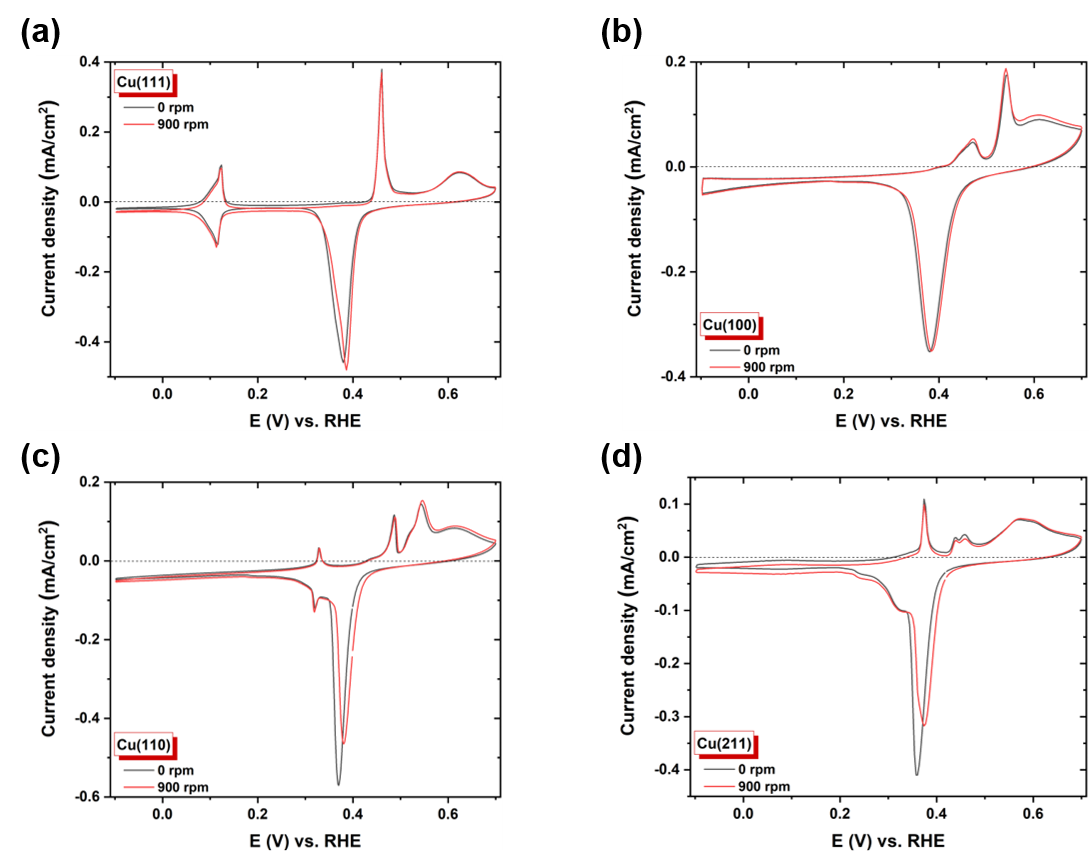}
    \caption{Polarization curves compared at 0 \& 900~rpm for (a) Cu(111), (b) Cu(100), (c) Cu(110), and (d) Cu(211) single crystals in Ar saturated 0.1~M KOH at a scan rate of 25~mV/s. These show that the oxidation currents observed in Figure 2 a \& b (main text), Figure 3 (main text), \cref{fig:Ar-CO-0rpm} (SI) and \cref{fig:Ar-CO-900rpm} (SI) are only due to CO oxidation without any artefact from the RDE measurements itself.}
    \label{fig:Ar-0 & 900rpm}
\end{figure*}

\section{Computational Methods}

DFT calculations were performed using Quantum Espresso \cite{espresso2009} with the BEEF-vdW exchange-correlation functional.\cite{BEEF2012} The plane-wave and density cutoffs used were 500 and 5000 eV respectively. A Fermi-smearing width of 0.1 eV was used, and the electronic structure was converged until the total energy difference was $<$ 10$^{-5}$ eV. Solvation effects were included via the self-consistent continuum solvation (SCCS) model as implemented in the Environ package.\cite{SCCS2012} We used the standard ``fitg03'' parameters ($\rho_\mathrm{(min)}$=0.0001, $\rho_\mathrm{(max)}$=0.005, $\alpha + \gamma$=11.5 dyn/cm).  Symmetric 3x4 slabs were used to simulate adsorbates on Cu(111) and Au(111) surfaces with a vacuum spacing of 12 $\AA$.  The Brillouin zone was sampled via a 4×3×1 Monkhorst-Pack grid.\cite{Monkhorst1976} Nonsymmetric slab models were used for nudged-elastic-band (NEB) calculations.\cite{NEB2000,NEB22000} The most stable adsorption sites for all adsorbates were identified through sampling of all symmetrically inequivalent adsorption sites as identified using the CatKit package.\cite{CATKIT2019} Barrier
calculations were performed using the NEB method and handled by AIDNEB.\cite{ASE2017,MLNEB2019} The barriers were computed with a minimal accuracy of the surrogate model of 25 meV and convergence of the forces on the climbing image of 25-50 meV. All transition states were confirmed to have a single imaginary frequency. Free energies were obtained following the ideal gas law for gas phase molecules and the harmonic oscillator model for the adsorbates, respectively. All energies were referenced to gas-phase H$_2$ (1 atm), CO (1 atm) and H$_2$O at the vapor pressure of liquid water (0.035 atm). To mitigate systematic DFT errors, we used corrections of 0.15 eV per C=O double bond and 0.1 eV for the H2(g) reference as suggested by  Christensen et al.\cite{Rune2015} Electrochemical reaction energetics were referenced using the Computational Hydrogen Electrode (CHE) formalism.\cite{CHE2004}.

\section{Computed reaction barriers on Cu(111)}

\begin{figure*}[hbt!]
    \centering
    \includegraphics[width=1\textwidth]{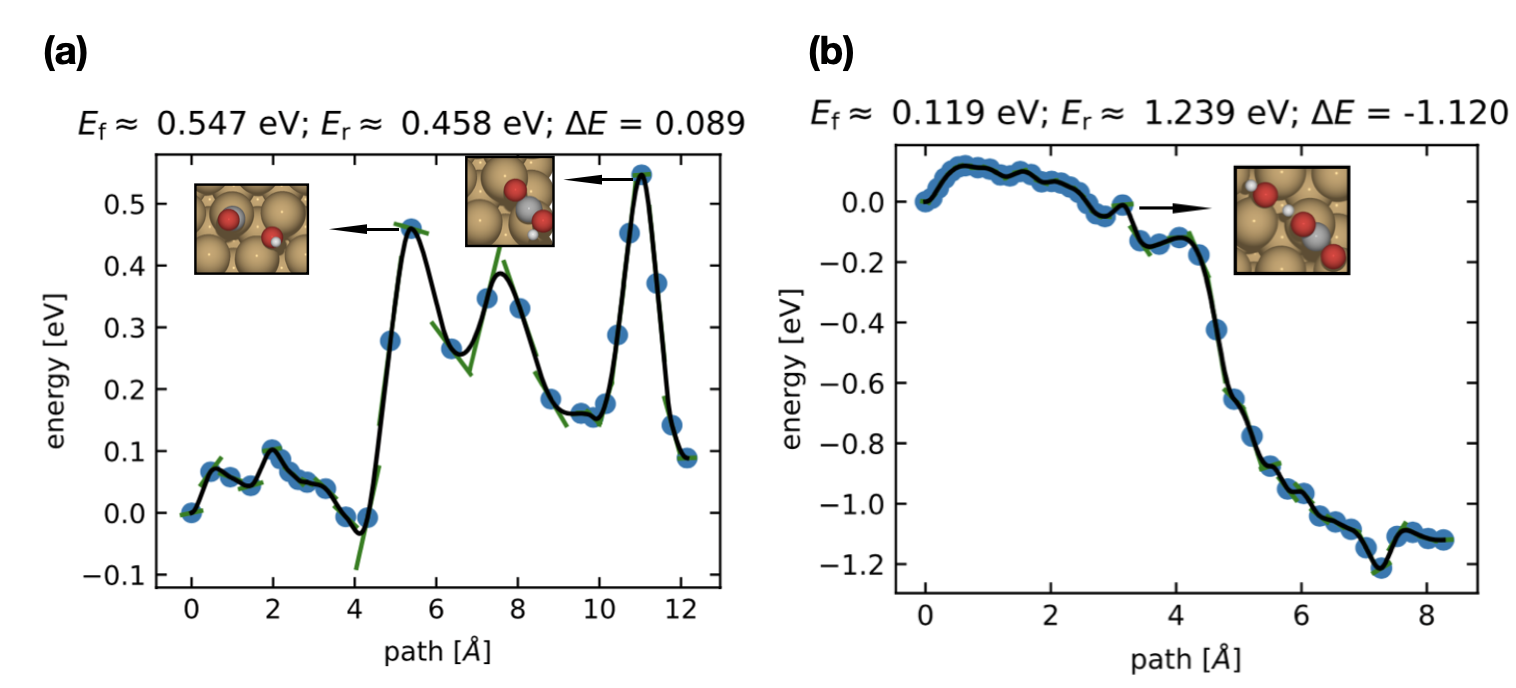}
    \caption{Potential energy profile from the converged NEB simulations for (a) the *CO-*OH coupling and (b) *COOH-*OH coupling on Cu(111). The blue points correspond to the energy of an individual NEB image and the green lines represent the forces (gradients). Transition states correspond to the image with the highest energy in the profile. Insets depict the atomic configuration of selected
images along the pathway. E$_f$, E$_r$, and $\Delta$E correspond to the forward barrier, reverse barrier and the energetic difference between the product and reactant states.}
    \label{fig:SI_NEB_pathway}
\end{figure*}

The energy profiles of the minimum energy pathway calculated using the NEB method for the *CO-*OH and *COOH-*OH coupling reactions on Cu(111) are provided in \cref{fig:SI_NEB_pathway}. Insets depict atomic configurations of key images along the NEB.

\section{Free energy profiles for e-COOR on Cu(111) and Au(111)}

\begin{figure*}[hbt!]
    \centering
    \includegraphics[width=1\textwidth]{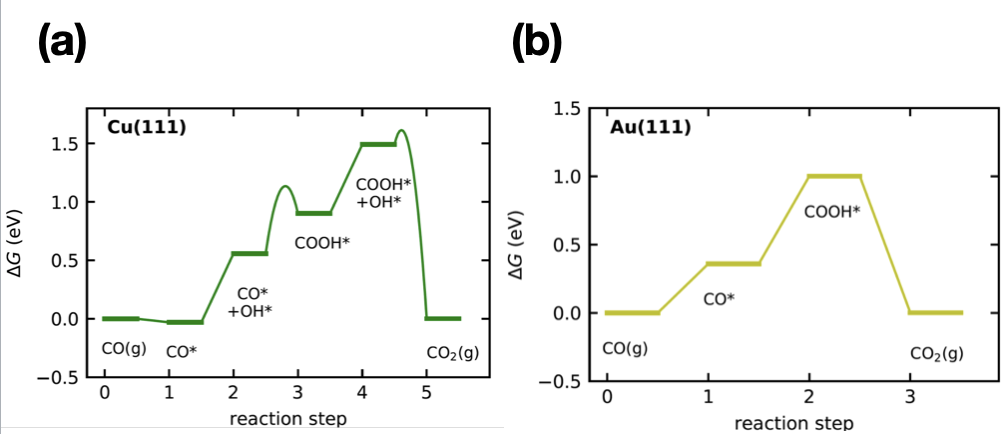}
    \caption{Free energy diagrams for the e-COOR on (a) Cu(111) and (b) Au(111) at the thermodynamic equilibrium potential (U = -0.22 V vs {RHE}) as predicted by DFT.}
    \label{fig:FEDs_Cu_Au}
\end{figure*}

\section{Approximation of the CO-OH$^{-}$ electrochemical barrier}

Decisive to the simulated polarization curves obtained using the microkinetic model are the barriers for the rate-determining step (RDS) in the reaction. For the example of CO oxidation on Au(111), where the RDS is clearly defined as the CO-OH$^{-}$ coupling step (see \cref{fig:FEDs_Cu_Au}), the barrier corresponds to an electrochemical reaction which cannot be reliably assessed \cite{Gauthier2019}. To this end, we estimate it using the data obtained from the RDE experiments on Au(111). In particular, we fit the simulated current density from  the microkinetic model to match the kinetic current density and Tafel slope obtained for Au(111) using the Levich analysis in Figure 2d (main text). The fitted barrier obtained for the electrochemical coupling of CO-OH$^{-}$ is ca. 0.6 eV. In addition, the charge transfer co-efficient, $\beta$ for this coupling step was assumed to be 0.5.

\begin{figure*}[hbt!]
    \centering
    \includegraphics[width=1\textwidth]{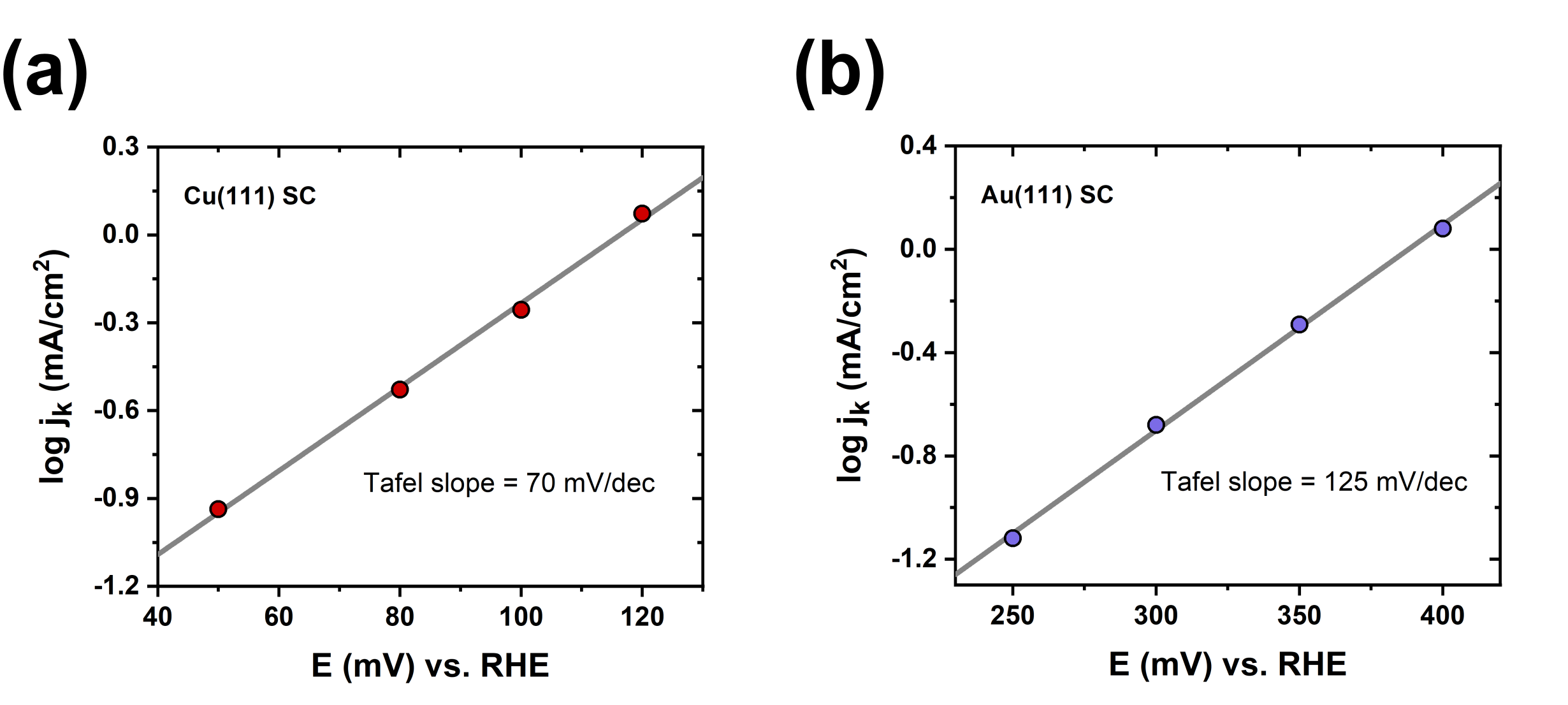}
    \caption{Tafel plot for (a) Cu(111) and (b) Au(111) using the Levich analysis in Figure 2c \& 2d (main text) respectively.}
    \label{fig:Tafel plot}
\end{figure*}

\begin{figure*}[hbt!]
    \centering
    \includegraphics[width=0.7\textwidth]{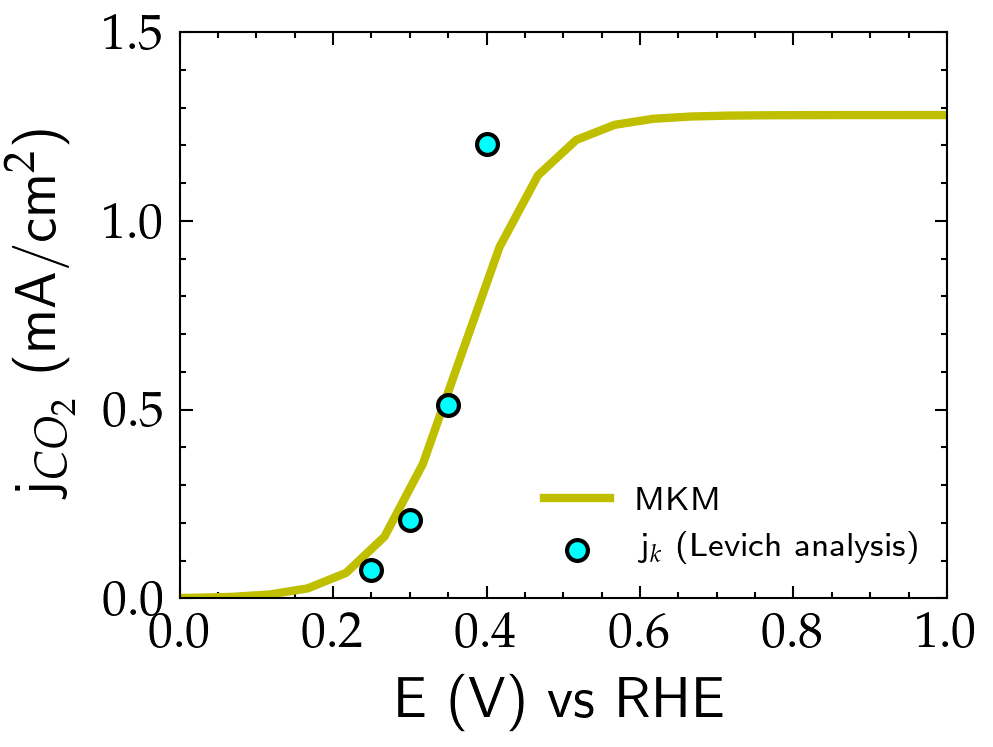}
    \caption{Simulated current density for e-COOR on Au(111) obtained using the microkinetic model (MKM) with an adjusted barrier for CO-OH$^{-}$ obtained by fitting to the kinetic current density (j$_k$) obtained using the Levich analysis on Au(111).}
    \label{fig:fitted_CO_OH}
\end{figure*}

\section{Details of the microkinetic model}

Microkinetic simulations were performed using the CatMAP package \cite{CATMAP2015} by using the DFT energetics as input. CatMAP solves the kinetic rate equations using the steady-state approximation in order to estimate species consumption, adsorbate coverages and production rates. The reactions shown in Eqns.\ (2)-(5) and (6)-(8) in the main text were used in the microkinetic model for e-COOR on Cu(111) and Au(111), respectively. A pre-exponential factor of 10$^{13}$ s$^{-1}$ as obtained from transition-state theory \cite{Norskov2014} was used for all the reaction steps. Current density towards CO$_2$ production (j$_{CO_2}$) was obtained from TOF$_{CO_2}$ using \cref{TOF2j} following Ref. \cite{heine2014}

 \begin{align} 
 	& \mathrm{j_{CO_2} = 2e\rho TOF_{CO_2}  } \label{TOF2j} \\
 	& \mathrm{e\rho = 80.3 \mu C/cm^2}
\end{align}

In order to account for the mass transport of the reactant CO species, we used Fick's first law to estimate its flux (J$_{CO}$) at the interface (cf. \cref{ficks}). The diffusion coefficient of CO was taken to be 20.3$\times$10$^{-10}$ m$^2$/s \cite{Cussler2009}. The concentration of dissolved CO (CO (aq)) is estimated from the Henry constant (1100 L.atm.mol$^{-1}$) \cite{Sander2015} and Henry's law is used to estimate the corresponding pressure of CO(g). 

 \begin{align} 
 	& \mathrm{J_{CO} = -D_{CO} \frac{dc}{dx}} \label{ficks}
\end{align}

A self-consistent cycle is run between CatMAP and the analytical diffusion model. The species consumption and product rates are estimated using the microkinetic rate equations solved by CatMAP.
The product rates obtained from the CatMAP model correspond to the turn-over-frequency (TOF) of
an individual catalytic site. The TOF is then converted to molar fluxes using a site-density
corresponding to the unit cell of the Cu(111) and Au(111) facets. These fluxes then serve as an input for the analytical model given by \cref{ficks} to obtain the CO concentration at the interface that is in turn used as the input in the CatMAP model. The microkinetic and transport models are then solved iteratively until a tolerance of 10$^{-2}$ is achieved.

\begin{figure*}[hbt!]
    \centering
    \includegraphics[width=1\textwidth]{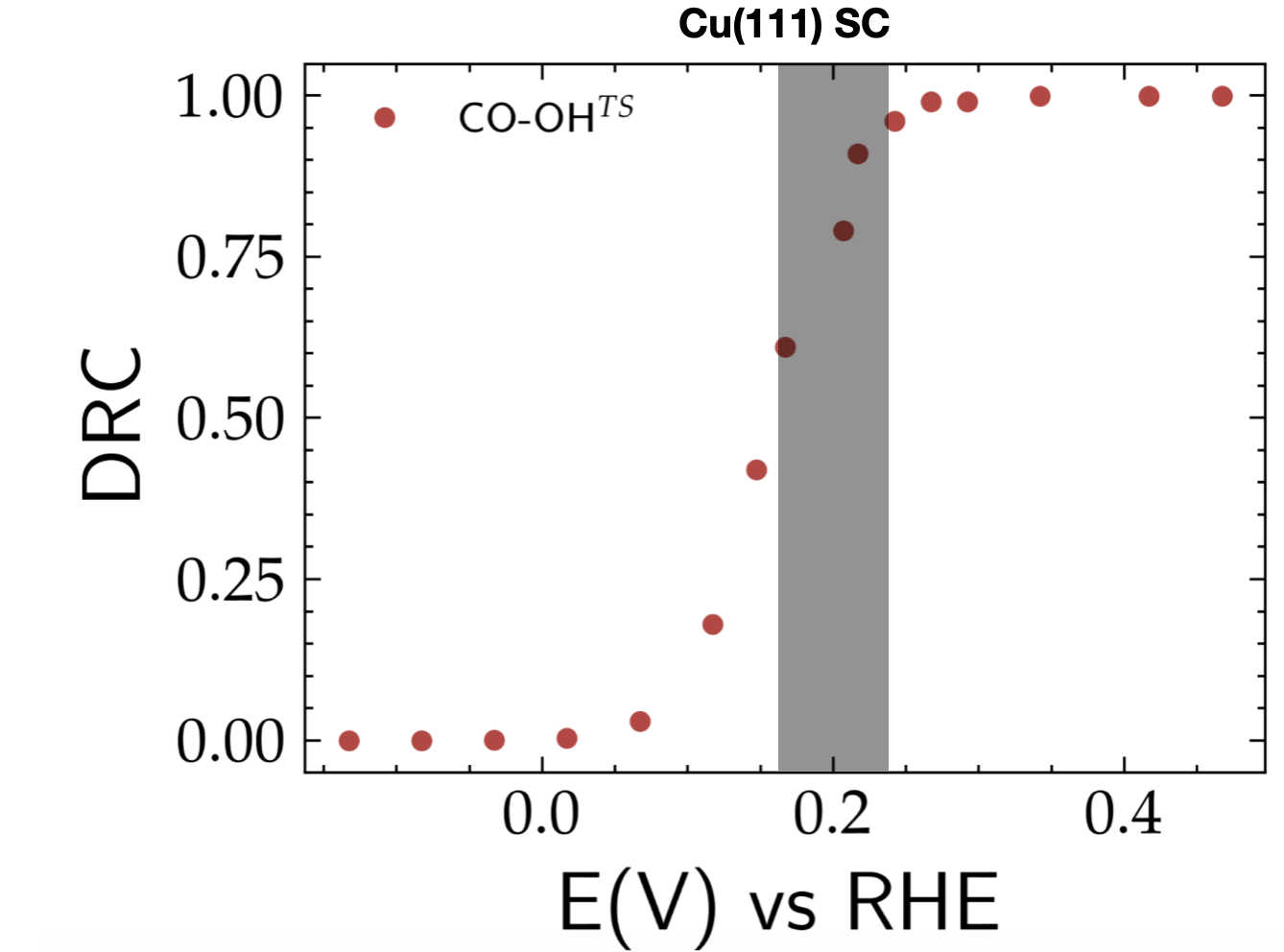}
    \caption{Simulated degree of rate control (DRC) obtained from the microkinetic simulations for the *CO-*OH transition state (TS) on the overall CO$_2$ production rate on the Cu(111) surface denoted as Cu(111) SC. A positive DRC value indicates that the intermediate has to be stabilized in order to increase the rate, and a DRC value of 1 corresponding to full rate control of the intermediate. The shaded region represents potentials close to the onset of e-COOR where the *CO-*OH$^{TS}$ dominates the overall rate of e-COOR.}
    \label{fig:SI_DRC}
\end{figure*}

\begin{figure*}[hbt!]
    \centering
    \includegraphics[width=1\textwidth]{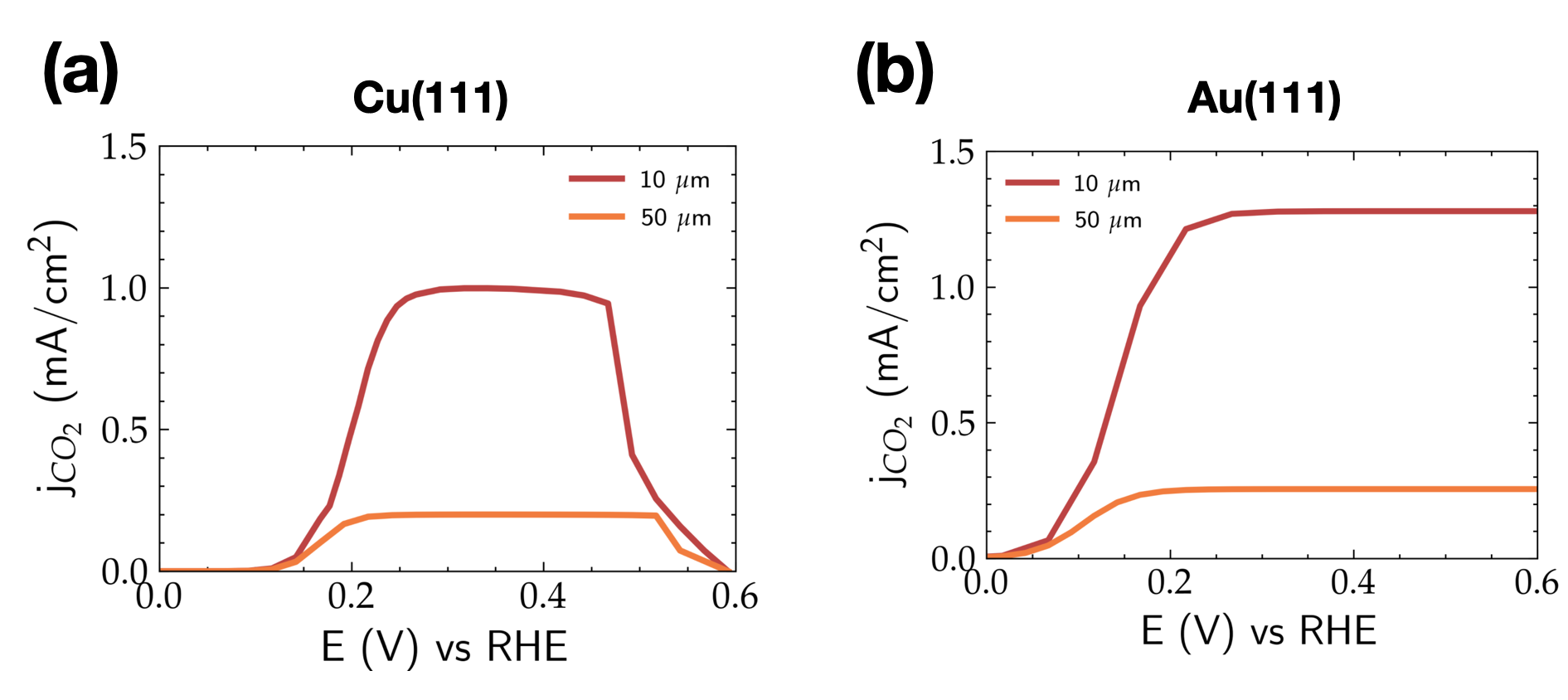}
    \caption{Sensitivity of the simulated CO$_2$ polarization curves to the boundary layer thickness for (a) Cu(111) and (b) Au(111) single crystals.}
    \label{fig:SI_BL_thickness}
\end{figure*}

\bibliography{RDE_ref}